\documentclass[seceq]{ptptex}

\usepackage[dvipdfm]{graphicx}
\usepackage{wrapft}

\notypesetlogo                       

\title{
Apparent Acceleration through Large-scale Inhomogeneities
}

\subtitle{\small Post-Friedmannian Effects of Inhomogeneities on the
  Luminosity  Distance}    

\author{
Masumi \textsc{Kasai}%
}

\inst{
Graduate School of Science and Technology, Hirosaki University\\ Hirosaki
036-8561, Japan
}

\recdate{March 14, 2007}

\abst{
  We re-analyze the observed magnitude-redshift relation of type Ia
  supernovae (SNe Ia) and examine the possibility that the apparent
  acceleration of the cosmic expansion is not caused by dark energy
  but is instead a consequence of the large-scale inhomogeneities in
  the universe.  We propose a method to phenomenologically describe  the
  effects of the large-scale inhomogeneities without relying  on
  the specific toy models of the inhomogeneous universe.  This method
  clearly illustrates how the post-Friedmannian effects of
  inhomogeneities, {\it i.e.} the effects due to the deviation from a 
  perfectly homogeneous and isotropic model, act as an effective
  cosmological constant in the magnitude-redshift relation of SNe Ia.
}

\begin{document}

\maketitle

\section{Introduction}

The Cosmological Principle, which states that our universe is
described by the homogeneous and isotropic
Friedmann-Lema\^{i}tre-Robertson-Walker (FLRW) model, in some averaged
sense, is a working hypothesis which has been widely accepted among
cosmologists.  
The present, past, and future evolution of the FLRW
model is determined by a few constant parameters, such as the Hubble
parameter, $H_0$, the matter density parameter, $\Omega_m$, and the
cosmological constant, $\Lambda$ (or the normalized parameter
$\Omega_{\Lambda}\equiv \Lambda/3 H_0^2$). The determination of the
cosmological parameters is one of the main purposes of observational
cosmology.

The recent observations of type Ia supernovae (SNe Ia) now strongly
suggest the acceleration of the cosmic
expansion.\cite{R98,P99,R04,As06}
As long as we employ perfectly
homogeneous and isotropic FLRW models, this requires dark energy, an
exotic energy component which accelerates the cosmic expansion with
its negative pressure.

Instead of introducing such a mysterious energy component, there have
been attempts to explain the apparent accelerated expansion of the
universe resulting from the large-scale inhomogeneities in the
universe. For example, Tomita,\cite{Tomi1,Tomi2,Tomi3,Tomi4,Tomi5}
using his local void model, and Iguchi et al. \cite{Igu}, using the
Lema\^{i}tre-Tolman-Bondi\cite{Lema,Tol,Bon} (LTB) model, studied the
possibility of explaining the observed magnitude-redshift ($m$-$z$)
relation of SNe Ia.  Moreover, recently Alnes et al. \cite{Al} have
concluded that not only the $m$-$z$ relation of SNe Ia but also the
position of the first peak in the cosmic microwave background (CMB)
anisotropy can be explained by the inhomogeneity in the LTB model.
However, these works depend specifically on simplified toy models.
Therefore, due to the lack of strong support for such toy models as
providing realistic descriptions of our universe, the study of
inhomogeneous effects in the universe is not the mainstream of
research in cosmology.

In this article, 
we re-analyze the observed $m$-$z$ relation of SNe Ia and point out
some theoretical possibilities of the inhomogeneity interpretation to
explain the apparent acceleration. 
Then, we propose a method to phenomenologically describe 
 the effects of the large-scale inhomogeneities in
the universe, without relying on  specific toy models.  This
method clearly illustrates how the post-Friedmannian effects of
inhomogeneities, {\it i.e.} the effects due to the deviation from
a perfectly homogeneous and isotropic FLRW model, act effectively as a
cosmological constant in the magnitude-redshift relation of SNe Ia.

\section{The magnitude-redshift relation of SNe Ia}

The apparent magnitude $m$ of a SN Ia of absolute magnitude $M$, at
redshift $z$, is 
\begin{equation}\label{eq:mz1}
  m = M + 5\log_{10} \frac{D_L(z)}{10\,\mbox{\small (pc)}}, 
\end{equation}
where $D_L(z)$ is the luminosity distance in units of parsecs. The
luminosity distance is obtained by solving the propagation of light
ray bundles through space-time.
In the FLRW universe, it is written
in the  form\cite{FFKT,Car}
\begin{eqnarray}\label{eq:DL}
  &&D_L(z) =
  \frac{c\,(1+z)}{{{H_0}}
    {\sqrt{1-{\Omega_m} -
        {\Omega_{\Lambda}}}}}  \nonumber\\
 &&\quad\times \sinh\left({\sqrt{1-{\Omega_m} - {\Omega_{\Lambda}}}} 
  \int_0^z \frac{ dz'}{\sqrt{(1+{\Omega_m}\, z')(1+z')^2 -
   z'\, (2+z') \, {\Omega_{\Lambda}}}} \right). 
\end{eqnarray}
The luminosity distance $D_L(z)$ is a slightly complicated function of
$z$ with three constant parameters, $H_0$, $\Omega_m$, and
$\Omega_{\Lambda}$.

As an illustration, we use the observed SNe Ia data presented in the
paper of Perlmutter et al.\cite{P99}. \ In total, 60 SNe Ia with
redshift in the range $0.014 \leq z \leq 0.830$ are listed in Tables 1
and 2 of Ref.~\citen{P99}. 
Because all of the SNe Ia have redshifts
satisfying $z < 1$, the luminosity distance $D_L(z)$ may be most
usefully expressed as a power series,
\begin{equation}
  \label{eq:DLtaylor}
  D_L(z) = \frac{c}{H_0} \left( z + {d_{2}}\, z^2 + {d_{3}}\, z^3 +
  \cdots
  \right), 
\end{equation}
where the expansion coefficients $d_2$ and $d_3$ are given by\cite{Ce} 
\begin{equation}
  {d_2} = \frac{1}{4}\left( 2 - \Omega_m + 2\,
    \Omega_{\Lambda}\right), 
\end{equation}
\begin{equation}
  {d_3} = \frac{1}{8} \left({\Omega_m}^2 + 4\, {\Omega_{\Lambda}}^2 -
    4\, \Omega_m  \,\Omega_{\Lambda} - 2\, \Omega_m - 4\,
    \Omega_{\Lambda}\right) . 
\end{equation}
Substituting  Eq.~(\ref{eq:DLtaylor}) into the $m$-$z$ relation
Eq.~(\ref{eq:mz1}) gives
\begin{eqnarray}\label{eq:mz2}
  m &=& M - 5 + 5\log_{10}D_L(z) \nonumber\\
 &=& {\cal M} + 5\log_{10} \left( z + {d_2}\, z^2 +
   {d_3}\,z^3\right), 
\end{eqnarray}
where ${\cal M} \equiv  M - 5 + 5\log_{10}{c}/{H_0}$. 
This quantity is often called
as the ``Hubble-constant-free absolute magnitude''\cite{P99} or the
``magnitude zero-point''.\cite{Ce} \
Note that  Taylor expansion up
to at least $O(z^3)$ is necessary  to determine the three parameters
$H_0$, $\Omega_m$, and $\Omega_{\Lambda}$ by  fitting to the data.
Once
the best fit coefficients  $d_2$ and $d_3$ are obtained from such a
fitting, we can calculate the cosmological parameters $\Omega_m$ and
$\Omega_{\Lambda}$ as follows:
\begin{eqnarray}\label{eq:getOm}
  {\Omega_m} &=& 2 \left(1 - {d_2}\right)\left(1-2\,{d_2}\right)
- 2\, {d_3}, \\
\label{eq:getOL}
  {\Omega_{\Lambda}} &=& {d_2} \left(2\,{d_2} - 1\right) - {d_3}. 
\end{eqnarray}

\begin{figure}[htbp]
  \centerline{
    \includegraphics{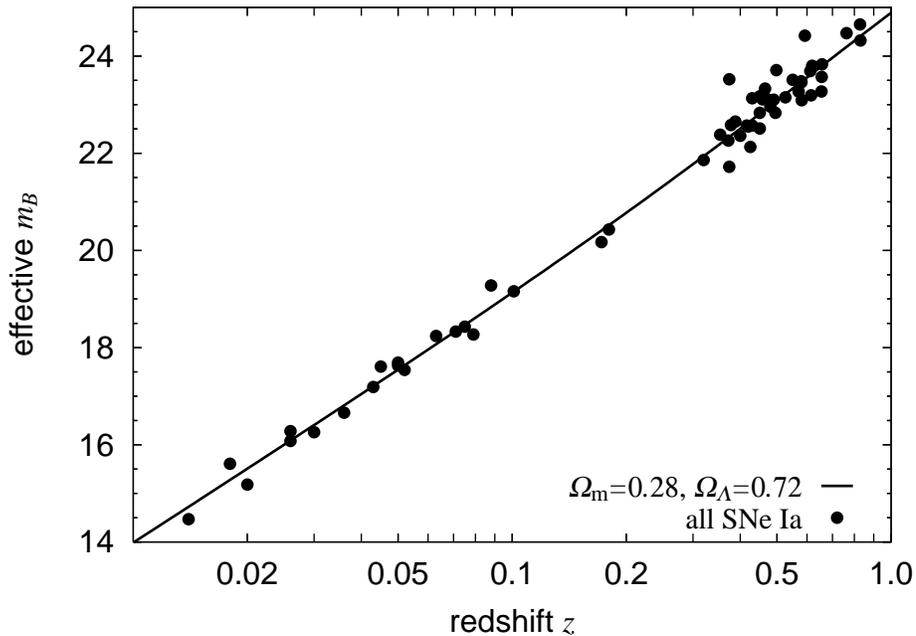}}
  \caption{Hubble diagram for all SNe Ia. These data are taken from
    Tables 1 and 2 of Perlmutter et al.\cite{P99}.\  The solid curve
    represents the best fit $m$-$z$ relation for a flat cosmology
    given by Perlmutter et al.\cite{P99}, with the parameter values
    $\Omega_m=0.28$ and $\Omega_{\Lambda}=0.72$. }
\end{figure}

In Fig.~1, we plot the Hubble diagram for all 60 SNe Ia.  Also plotted
there is the best fit $m$-$z$ curve for a flat cosmology with the
parameter values $\Omega_m=0.28$ and $\Omega_{\Lambda} = 0.72$, which
was determined by Perlmutter et al.\cite{P99}.

\section{Inhomogeneous interpretation?}

In order to formulate an alternative interpretation of the $m$-$z$
relation without dark energy, it is instructive to divide the whole
SNe Ia data set into two parts.  For the sake of convenience, we
define SNe Ia with redshifts satisfying $z<0.2$ as low-redshift
(low-$z$) SNe Ia, and those with redshifts satisfying $z>0.3$ as
high-redshift (high-$z$) SNe Ia.  Then, the low-$z$ data set consists
of 20 SNe Ia with redshifts in the range $0.014 < z < 0.18$, and the
high-$z$ data set consists of 40 SNe Ia with redshifts in the range
$0.320 < z < 0.830$.  No SNe Ia are in the range $ 0.2 \leq z \leq
0.3$.

\begin{figure}[htbp] 
\centerline{\includegraphics{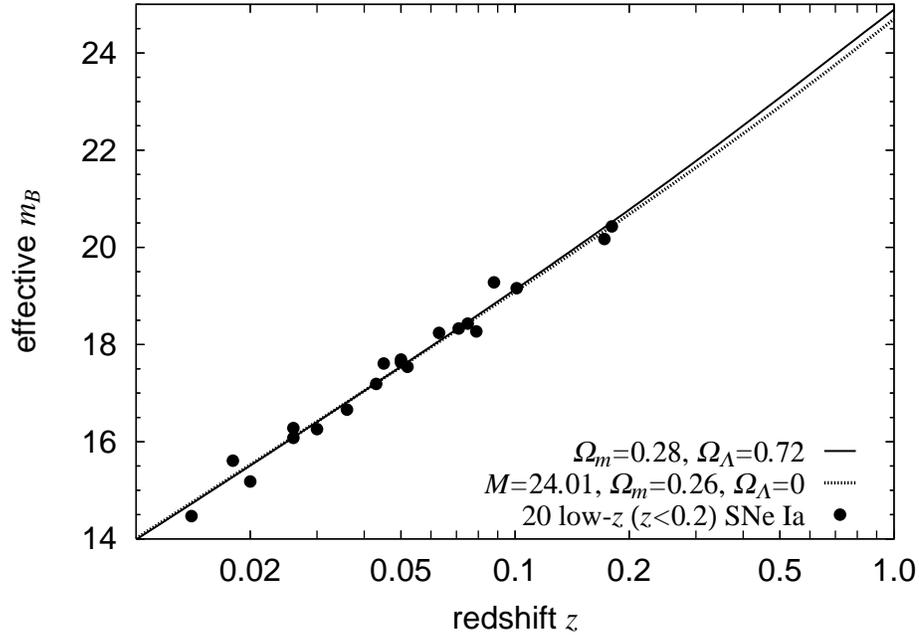}}
\caption{Hubble diagram for the 20 low-redshift ($z<0.2$) SNe Ia data
  set and the best fit $m$-$z$ curve of a {zero-$\Lambda$} cosmology
  with the parameter values ${\cal M}=24.01$ and $\Omega_m=0.26$. }
\end{figure}

\begin{figure}[htbp]
  \centerline{\includegraphics{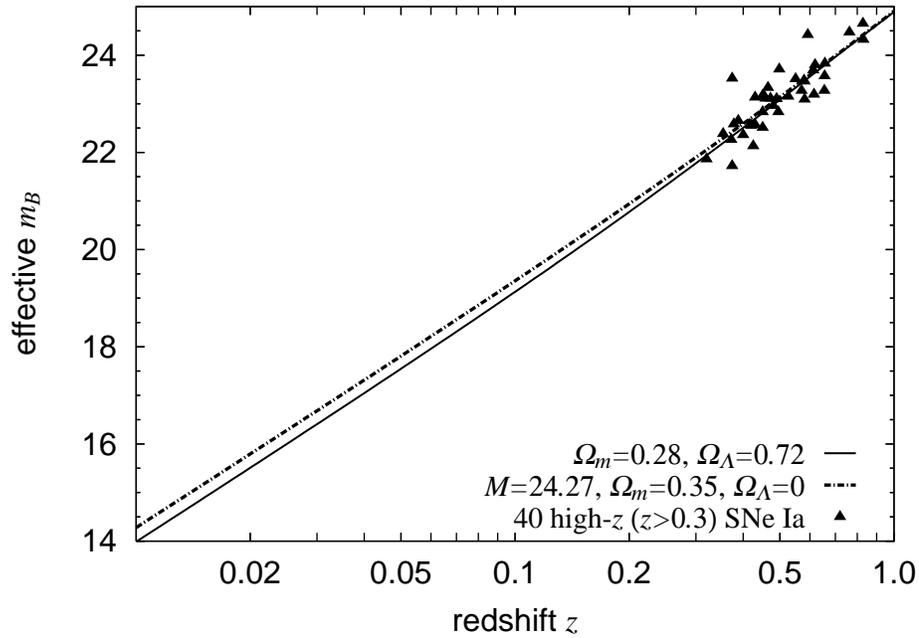}}
  \caption{Hubble diagram for the 40 high-redshift ($z>0.3$) SNe Ia
    data set and the best fit $m$-$z$ curve for a {zero-$\Lambda$}
    cosmology with the parameter values ${\cal M}=24.27$ and
    $\Omega_m=0.35$.  }
\end{figure} 

In this paper, we use an implementation of the nonlinear least-square
methods of the Levenberg-Marquardt algorithm\cite{Lev,Mar} to fit the
$m$-$z$ relation Eq.~(\ref{eq:mz2}) to each data set.

Figure~2 displays only the low-$z$ SNe Ia data set, along with the
best fit $m$-$z$ curve of a {zero-$\Lambda$} cosmology with the
parameter values ${\cal M}=24.01$ and $\Omega_m=0.26$.  Since the
cosmological constant $\Lambda$ does not play an important role for
$z\ll 1$, it is evident that the {zero-$\Lambda$} cosmology can fit
the low-$z$ data set fairly well.

Figure~3 displays only the high-$z$ SNe Ia data set, along with the
best fit $m$-$z$ curve of a {zero-$\Lambda$} cosmology with the
parameter values ${\cal M}=24.27$ and $\Omega_m=0.35$.  It is
interesting that a {zero-$\Lambda$} cosmology can also accurately fit
the high-$z$ SNe Ia data set, although the values of the cosmological
parameters ${\cal M}$ and $\Omega_m$ in that case are different from
the best fit values for the low-$z$ SNe Ia.  It should also be noted
that the high-$z$ best fit $m$-$z$ curve (${\cal M}=24.27,
\Omega_m=0.35, \Omega_{\Lambda}=0$) behaves very similarly to that of
the flat cosmology with $\Omega_m=0.28$ and $\Omega_{\Lambda}=0.72$
obtained by Perlmutter et al.\cite{P99} in the range $0.3 < z < 1$.

\begin{figure}[htbp]
  \centerline{
    \includegraphics{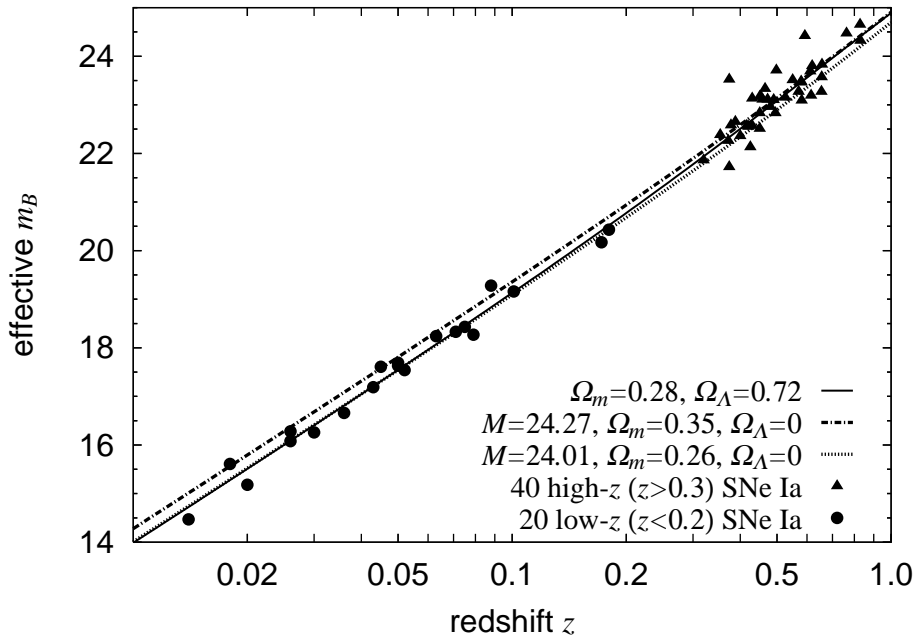}}
  \caption{Hubble diagram for the 20 low-redshift ($z<0.2$) SNe Ia and
    the 40 high-redshift ($z>0.3$) SNe Ia. Also plotted are the best
    fit $m$-$z$ curves of a {zero-$\Lambda$} cosmology for each data
    set and the best fit flat cosmology with $\Omega_m=0.28$ and
    $\Omega_{\Lambda}=0.72$ for the entire SNe Ia data set obtained by
    Perlmutter et al.\cite{P99} as a reference. }
\end{figure} 

Figure~4 summarizes all of the above results.  There it is seen that
an open {zero-$\Lambda$} FLRW model with the parameter values ${\cal
  M}=24.01$ and $\Omega_m=0.26$ can fit the low-$z$ SNe Ia data,
whereas another open {zero-$\Lambda$} FLRW model with ${\cal M}=24.27$
and $\Omega_m=0.35$ can fit the high-$z$ data.  No cosmological
constant nor dark energy is necessary to fit either part.  However, a
positive cosmological constant ($\Omega_{\Lambda}=0.72$) is necessary
to fit the entire SNe Ia data set with a single flat FLRW cosmology.
It is evident from Fig.~4 that this conclusion simply comes from the
asymptotic behavior of the theoretical curve with $\Omega_m=0.28$ and
$\Omega_{\Lambda}=0.72$ and does not depend on the particular choices
of the SNe Ia data sets.

Because ${\cal M} = M - 5 + 5\log_{10}c/H_0$, where the Hubble
distance $c/H_0$ is in units of parsecs, the result that different
values of $\cal M$ fit different redshift data sets implies the
following possibilities:
\begin{enumerate}
\item  The absolute magnitude $M$ of the high-$z$ SNe Ia is
  systematically different from that of the low-$z$ SNe Ia.
\item The speed of light $c$ is different in different redshift
  regions.
\item $H_0$ in the high-$z$ region is slightly different from
  that in the low-$z$ region.
\end{enumerate}

Let us examine the third possibility, namely, the inhomogeneity
interpretation.  If the difference between ${\cal M}\mbox{\scriptsize
  (low-$z$)}=24.01$ and ${\cal M}\mbox{\scriptsize (high-$z$)} =24.27$
is due to the inhomogeneity of the Hubble parameter $H_0$, we can
estimate the following ratio representing the inhomogeneity in $H_0$
between the two different redshift regions as
\begin{equation}
  24.01 - 24.27 =
  5\log_{10}\frac{H_0\mbox{\scriptsize(high-$z$)}}
  {H_0\mbox{\scriptsize(low-$z$)}} .    
\end{equation}
This yields $H_0\mbox{\scriptsize(high-$z$)} =
0.89\,H_0\mbox{\scriptsize(low-$z$)}$.  An inhomogeneity in which the
value of $H_0$ in the high-$z$ region is $11$\% smaller than that in
the low-$z$ region may be sufficient to explain the observed $m$-$z$
relation for SNe Ia, without the need to introduce a cosmological
constant or dark energy.

\section{Effects of large-scale inhomogeneities on the luminosity
  distance} 

In the previous section, we pointed out the interesting possibility
that it may be possible to account for the observed $m$-$z$ relation
for SNe Ia by the large-scale inhomogeneities in the universe, without
introducing a cosmological constant or dark energy.  In order to
examine this possibility further, we need to study the propagation of
light ray bundles through the inhomogeneous universe and obtain the
luminosity distance $D_L(z)$ as a function of the redshift $z$.  Since
the actual space-time inhomogeneities of the present universe are not
known in detail, the usual approach  employs
simplified toy models of the inhomogeneous universe.  For example,
Tomita\cite{Tomi1,Tomi2,Tomi3,Tomi4,Tomi5} used a local void model,
and Iguchi et al.\cite{Igu} used the LTB model.

We take another approach.  In this section, we propose a method to
phenomenologically describe the effects of the large-scale
inhomogeneities on the luminosity distance, without relying on
specific toy models of the inhomogeneous universe.  This method
clearly illustrates how the ``post-Friedmannian'' effects of
inhomogeneities, {\it i.e.} the effects due to the deviation from a
perfectly homogeneous and isotropic FLRW model, act effectively as a
cosmological constant in the magnitude-redshift relation of SNe Ia.

In the perfectly homogeneous FLRW models, $H_0$ denotes the expansion
rate at the present time, $t_0$, and it is constant over the entire
$t=t_0$ hypersurface, due to the perfect spatial homogeneity.  In
inhomogeneous universes, however, the expansion rate is naturally
dependent on the spatial positions. Therefore, $H_0$ is not constant
and may depend on $z$:
\begin{equation}
H_0\Rightarrow H_0(z). 
\end{equation}
Analogously, the density parameter $\Omega_m$ may also depend on $z$:
\begin{equation}
\Omega_m\Rightarrow\Omega_m(z) . 
\end{equation}

In general inhomogeneous universes, these cosmological parameters may
also depend on the angular direction due to spatial anisotropies.
Already in 1966, Kristian and Sachs\cite{KS} emphasized the importance
of observing angular variations in the various cosmological effects.
Kasai and Sasaki\cite{KaSa} and Kasai\cite{Kasa1988} derived formulae
for cosmological observations in a linearly perturbed FLRW model in
gauge-invariant manner and found the existence of a quadrupole
anisotropy of the Hubble parameter $H_0$, which is directly
proportional to the gauge-invariant scalar potential.  They found that
``the perturbed space-time behaves as a Friedmann-like universe with
the direction-dependent $H_0$ and $q_0$''.\cite{Kasa1988} In this
paper, however, we concentrate on the $z$ dependence of the
cosmological observables.  Future investigations will include
consideration of the angular dependences of $H_0$ and $\Omega_m$.
 
In the region $z < 1$, the cosmological parameters can be expressed in
the power series forms
\begin{eqnarray}\label{eq:H0z}
  H_0(z) &=& \bar{H}_0 \left( 1 + h_1\,z + h_2\,z^2 + \cdots\right),\\
  \Omega_m(z) &=& \bar{\Omega}_m \left( 1 + \omega_1\,z + \omega_2\,z^2 +
    \cdots\right),   \label{eq:Omz}
\end{eqnarray}
where $\bar{H}_0=H_0(z=0)$ and $\bar{\Omega}_m=\Omega_m(z=0)$, and the
expansion coefficients $h_1, h_2, \dots$, $\omega_1, \omega_2,\dots$
represent the ``post-Friedmannian'' corrections due to spatial
inhomogeneities.  The models reduce to the FLRW if and only if all of
these coefficients vanish.  Under the assumption that $H_0(z)$ and
$\Omega_m(z)$ are slowly varying functions of $z$, we substitute them
for $H_0$ and $\Omega_m$ in Eq.~(\ref{eq:DL}) and obtain the following
power series formula:
\begin{equation}\label{eq:DLz3}
  D_L(z) = \frac{c}{\bar{H}_0} (z + \tilde{d}_2\,z^2 +
  \tilde{d}_3\,z^3 + \cdots), 
\end{equation}
where the expansion coefficients are now
\begin{equation}
  \tilde{d}_2 = \frac{1}{4}\left( 2 - \bar{\Omega}_m + 2\, {\Omega}_{\Lambda}
  \right) - h_1, 
\end{equation}
\begin{eqnarray}
  {\tilde{d}_3} &=& \frac{1}{8} \Bigl({\bar{\Omega}_m}^2 + 4\,
  {\Omega_{\Lambda}}^2 - 4\, \bar{\Omega}_m 
  \,\Omega_{\Lambda} - 2\, \bar{\Omega}_m - 4\, \Omega_{\Lambda}
  \Bigr)\nonumber \\
  &&
  + \frac{1}{4}\left\{4(h_1)^2-4 h_1 - 4 h_2 -
    h_1 \,(2\Omega_{\Lambda}-\bar{\Omega}_m-2) -
    \frac{2}{3}\omega_1\,\bar{\Omega}_m \right\}. 
\end{eqnarray}

Now we can illustrate how the ``post-Friedmannian'' corrections of the
spatial inhomogeneities act effectively as a cosmological constant.
Suppose that astronomers obtain the best fit parameters $\tilde{d}_2$
and $\tilde{d}_3$ by fitting the luminosity distance formula
Eq.~(\ref{eq:DLz3}) to the observed $m$-$z$ data of SNe Ia.  If they
assume that the universe is homogeneous and isotropic, they will
simply use Eqs.~(\ref{eq:getOm}) and (\ref{eq:getOL}) to calculate the
cosmological parameters.  Then, even if the true value of the
cosmological constant is zero, they will obtain the following
effective value for the cosmological constant from
Eq.~(\ref{eq:getOL}):
\begin{eqnarray}
\Omega_{\Lambda}^{\mbox{\scriptsize eff}} &\equiv&
 {\tilde{d}_2}\, \left(
   2\,{\tilde{d}_2} - 1\right) - {\tilde{d}_3} \nonumber\\
 &=& \frac{1}{4} \left\{ 3 {h_1} \bar{\Omega}_m 
 + \frac{2}{3} {\omega_1} \bar{\Omega}_m - 2
 {h_1}
 + 4 \left({h_1}\right)^2 + 4 { h_2} \right\}\label{eq:OLeff}
\end{eqnarray}
This clearly shows that the ``post-Friedmannian'' correction terms
$h_1, h_2$, and $\omega_1$ together act effectively as a cosmological
constant.  The effective value
$\Omega_{\Lambda}^{\mbox{\scriptsize eff}}$
is unrelated to the true value of the cosmological
constant. It simply results from the erroneous assumption that the
universe is perfectly homogeneous and that $H_0$ and $\Omega_m$ are
constant on the $t=t_0$ hypersurface.

In the same way, we  also obtain the following effective value for
the density parameter from Eq.~(\ref{eq:getOm}): 
\begin{eqnarray}
  \Omega_m^{{\mbox{\scriptsize eff}}} &\equiv&
     2 \left(1 - {\tilde{d}_2}\right)\left(1-2\,{\tilde{d}_2}\right)
     - 2\, {\tilde{d}_3} \nonumber\\
     &=& \left(1+\frac{3}{2}h_1 + \frac{1}{3}\omega_1\right)
     \bar\Omega_m
     + 3 h_1 + 2(h_1)^2 + 2h_2. 
\end{eqnarray}

Following the procedure described in \S 2., we obtain the best fit
values for the cosmological parameters as
$\Omega_m^{{\mbox{\scriptsize eff}}} = 0.28$ and
$\Omega_{\Lambda}^{{\mbox{\scriptsize eff}}}=0.72$. Unfortunately,
however, these do not completely determine the inhomogeneities of the
actual universe.  This is simply because the two cosmological
parameters $\Omega_m^{{\mbox{\scriptsize eff}}}$ and
$\Omega_{\Lambda}^{{\mbox{\scriptsize eff}}}$ are functions of four
parameters $\bar\Omega_m, h_1, h_2$, and $\omega_1$ (or possibly five,
including $\Omega_{\Lambda}$).  The data fitting of the $m$-$z$
relation itself only yields constraints on some sets of the
post-Friedmannian parameters, but it does not determine completely the
values of each parameter independently.  For this reason, it is highly
desirable to incorporate other independent observations, such as CMB
data, gravitational lensing data, and so on, in order to determine the
extent to which our universe is homogeneous or inhomogeneous.

\section{Summary}

We have re-analyzed the observed $m$-$z$ relation of SNe Ia proposed
by Perlmutter et al.\cite{P99} and have examined the possibility that
the apparent acceleration of the cosmic expansion is a consequence of
large-scale inhomogeneities in the universe.  As previously found by
Perlmutter et al.\cite{P99}, a positive cosmological constant is
necessary to fit the whole data set, consisting of 60 SNe Ia in the
redshift range $0.014 \leq z \leq 0.830$, with a single FLRW model.
They obtained the best fit values $\Omega_m=0.28$ and
$\Omega_{\Lambda}=0.72$ for a flat cosmology.

In order to examine the feasibility of the inhomogeneity
interpretation, we divided the SNe Ia data into two parts, low-$z$ and
high-$z$ data sets.  The low-$z$ ($z<0.2$) data set consists of 20 SNe
Ia in the redshift range $0.014 \leq z \leq 0.18$, and the high-$z$
($z>0.3$) data set consists of 40 SNe Ia in the range $0.320 \leq z
\leq 0.830$.  We were able to fit the low-$z$ and high-$z$ data sets,
respectively, with two {zero-$\Lambda$} FLRW cosmologies.  The best
fit parameters are ${\cal M}=24.01$ and $\Omega_m=0.26$ for the
low-$z$ data set and ${\cal M}=24.27$ $\Omega_m=0.35$ for the high-$z$
data set.  The difference between the values of $\cal M$ implies that
the Hubble parameter $H_0$ in the high-$z$ region is 11\% smaller than
that in the low-$z$ region.  This indicates the possibility of the
cosmological parameters $H_0$ and $\Omega_m$ being dependent on the
redshift $z$.  It also suggests that the nearby low-$z$ region is a
local void, {\it i.e.} a less dense and more rapidly expanding than
the outer high-$z$ region. This local void interpretation is
consistent with the results of previous works employing inhomogeneous
toy models.\cite{Tomi1,Tomi2,Tomi3,Tomi4,Tomi5}\cite{Igu,Al}

Inspired by the above results from the data fittings, we proposed a
method to phenomenologically describe the effects of the large-scale
inhomogeneities, without relying on specific toy models of the
inhomogeneous universe.  In general inhomogeneous universes, the
Hubble expansion rate $H_0$ and the density parameter $\Omega_m$ are
naturally dependent on the spatial coordinates, and therefore the
redshift $z$.  We expanded $H_0(z)$ and $\Omega_m(z)$ into power
series with respect to $z$ and obtained a luminosity distance formula
with corrections due to the large-scale inhomogeneities.  This
distance formula clearly illustrates how the corrections resulting
from the large-scale inhomogeneities act effectively as a cosmological
constant.

In this paper, we re-analyzed only the SNe Ia data presented by
Perlmutter et al.\cite{P99} Other data sets, including those with
$z>1$ SNe Ia, should also be examined.  It is also noted that the
$m$-$z$ relation of SNe Ia itself does not completely determine the
inhomogeneities of the actual universe.  The data fitting of the
$m$-$z$ relation itself only gives constraints on some sets of the
post-Friedmannian parameters.  It does not completely determine the
values of each parameter independently.  For this reason, it is highly
desirable to incorporate other independent observations, such as CMB
data, gravitational lensing data, and so on, in order to determine the
extent to which our universe is homogeneous or inhomogeneous.


%

\end{document}